\documentclass[12pt]{article}
\usepackage[english]{babel}
\usepackage{amsmath}
\usepackage{times}
\usepackage[body={6.5in,9.5in}]{geometry}

\usepackage{amsmath}
\usepackage{url}
\usepackage{graphicx}
\usepackage{subfigure}
\usepackage{color}

\title{Theory and applications
of the Vlasov equation \footnote{Accepted for publication in EPJD}
 }

\author{{\it F. Pegoraro}$^{1}$, {\it F. Califano}$^{1}$, {\it G. Manfredi}$^{2}$, {\it P.J. Morrison}$^{3}$\\
{\small $^{1}$ Department of Physics ``Enrico Fermi'', University of Pisa, Italy}\\
{\small  $^{2}$ Centre National de la Recherche Scientifique}\\ {\small  Institut de Physique
et Chimie des Mat\'eriaux de Strasbourg, France}\\
{\small $^{3}$  Physics Department and Institute for Fusion Studies,}\\ {\small  the University of Texas at Austin, USA}}

\date{}

\begin{document}

\maketitle

\abstract{Forty  articles have been recently published in EPJD as contributions to  the  topical issue ``Theory and applications
of the Vlasov equation".  The aim of this topical issue was 
to provide a forum for the presentation of a broad variety of scientific
results involving
the Vlasov equation.  \,\, In this editorial, after some introductory notes,   a brief  account  is given of the main points addressed in these papers and of the perspectives they open.

\maketitle
\section{Introduction}\label{S1}

Referring to the system of equations that now goes under the name of the Vlasov-Poisson system, in his celebrated 1946 paper \cite{Landau} `` On the vibrations of an electronic plasma''  Lev Landau wrote a very sharp remark: ``These equations were used by A.A. Vlasov for investigation of the vibrations of plasma. However most of his results are incorrect.''\\
Strange indeed is the fate of the name of this fundamental equation for plasma physics that was neither discovered by, nor correctly solved by, the man whose name it bears, although Vlasov correctly recognized \cite{Vlasov} that ``for a system of charged particles the kinetic equation method which considers only binary interactions - interactions through collisions - is an approximation which is strictly speaking inadequate, so that in the theory of such systems an essential role must be played by the interaction forces, particularly at large distances and, hence, a system of charged particles is, in essence, not a gas but a distinctive system coupled by long-range forces''. In fact this equation was used  in the context of star dynamics, we would now say of galactic dynamics, one hundred years ago by J. Jeans  \cite{Jeans} and  is still  generally known in the astrophysics community \cite{Henon}  as the Collisionless Boltzmann equation.  This latter denomination  however does not do  real  justice to the full physics content of this equation as, presenting it as a limiting case of  the Boltzmann equation,  it does not emphasize   the novelty introduced  by  its intrinsic nonlinearity that arises from the collective long-range gravitational, or electromagnetic,  multibody interactions.\\
At any rate ``the Vlasov equation'' is presently  the most commonly used  trademark of the equation that describes the phase-space classical dynamics (as opposed to quantum dynamics) of multibody systems where the collective (mean field) interaction due to the long-range forces between its elementary constituents overwhelms the effects of their discrete binary interactions, generally known as Coulomb collisions. Depending on the system under consideration the long range interaction may be  electromagnetic, in which case the Vlasov equation must be solved together with Maxwell's equations, or gravitational,  in which case the Vlasov equation must be solved together with Newton's  or  Einstein's equations. 
In all these systems  the Vlasov equation allows us to determine the selfconsistent time evolution of the source terms of the electromagnetic or  of the gravitational fields.
But the solution of the Vlasov equation gives us much more than these source terms as it gives us the time evolution of the full distribution function  in phase space of the elementary constituents, particles from now on, of the system. This opens up a new realm of phenomena that are not included in the corresponding set of fluid equations which, at least  in principle,  would also make it possible to determine a time evolution for the source terms of the gravitational and electromagnetic fields.  Of major importance is the phenomenon of phase space mixing of the particle distribution function that is at the basis of the collisionless damping of the  perturbations of the particle density and of the electric field discussed by Landau in  his 1946 paper.  The non dissipative nature of this  damping represented a surprising novelty for the plasma community and took some time to become accepted until  its experimental  verification  by  Malmberg  and  Wharton \cite{Malmberg} nearly  twenty  years  after  Landau's  original  paper. 
The mathematical subtleties  involved in the description of  this phenomenon  are made evident in the recent work, which was awarded the 2010 Fields medal,  by the mathematician  C. Villani, see e.g., his recent  tutorial presentation in Ref. \cite{Villani}.

The physics applications of the Vlasov equation range from  magnetically confined plasmas for thermonuclear research to space plasmas in planetary magnetospheres and in stellar winds, to relativistic electromagnetic plasmas  either produced in the interaction of ultraintense laser pulses with matter or present  in the  high energy density environment around   compact astrophysical objects. Widely different time and spatial scales characterize these systems but they all share the mark of their nonlinear collective behaviour insofar as their macroscopic properties are determined by the excitations of waves and instabilities in conditions that are generally  very far  from even approximate  local  thermodynamic equilibrium.   The Vlasov equation applies also to the gravitational dynamics of the stellar component of galaxies, where the stars are the particles of the system. In this case  the typical characteristic collective dynamical time scale is approximately of the order of one hundredth of the age of the Universe, so that, in general, galactic dynamicists are less interested in the full long term development of the Vlasov dynamics. This   includes  phenomena such as the so called violent relaxation  \cite{lynden} of a collapsing stellar system that also occurs on the dynamical time scale.

When solving   the Vlasov equation  we must address  some of the most advanced problems in the physics of dynamical systems, in the  stability of Hamiltonian systems with infinite degrees of freedom \cite{Morrison}, in  kinetic turbulence and Hamiltonian chaos.  Recently, the  development of accurate numerical schemes that preserve the symplectic nature of the Vlasov dynamics and the availability of increasingly powerful supercomputers  have made it possible to explore solutions of the Vlasov equation well beyond the reach of analytical methods.  These  developments are making  the numerical integration of the  Vlasov equation an increasing effective and practical  investigation tool even in high dimensionality  fully nonlinear plasma regimes, thus  limiting our dependence on reduced models, such as e.g., fluid-type equations.  Together with the ability to    produce  such collisionless plasma regimes in the laboratory,  these numerical investigation tools have brought to light  new physics phenomena, e.g., regarding the mechanisms of acceleration of charged particles to ultra high energies in relativistic plasmas. Such  collective mechanisms are  not easily described  within a reductionist approach in terms of elementary single particle processes and their identification is extremely important for  understanding  the rich workings  of our  Universe.

It must  finally be mentioned that presently there are  active research  lines  that are  extending the scope of the Vlasov equation to new physical regimes in a variety of directions, including quantum  and incoherent high frequency radiation effects, see e.g., \cite{tamb,square},  electron-positron pair production \cite{SSB} up to the semiclassical treatment of non-Abelian field theories in quark-gluon plasmas \cite{Ipp}.  Most of these extensions are motivated  by their applicability to high energy astrophysics \cite{SVB}.  However in some of these new developments dissipative type of effects, not present in the original collisionless  Vlasov equation, are reintroduced due to high energy processes not present at lower plasma energies.

\section{Topical issue} 

Partly on the occasion of  a conference\footnote{VLASOVIA 2013, \url{http://vlasovia2013.event.univ-lorraine.fr/vlasov_home.php}}  held in Nancy in November   2013, a call was launched  to all interested  researchers worldwide  for papers to be published as  part of a topical issue in EPJD.  This call for papers on the  ``Theory and applications of the Vlasov equation'' attracted a large number of very valuable contributions. \, \, The full  list of the forty papers published, most of them  in 2014 (but not in a single issue of EPJD),  as contributions  to this topical issue are listed on the EPJD web site  \url{ http://epjd.epj.org/component/toc/?task=topic&id=274}.

These papers cover  a  comprehensive set of the  different  present-day  lines of research in the study of the Vlasov  equation.  
These range  from  revisiting  the  foundations  of the Vlasov equation, its relationship with the  single particle dynamics, its Hamiltonian properties,  the preservation  of such properties  in the reduction of the full Vlasov  description in fluid-type models, in Landau fluid or gyrokinetic models  and finally  in  numerical implementations,
to the description of new, time independent solutions or to the use of special classes of solutions such as the  so-called Water Bag distributions, to the study  of coherent  nonlinear  
solutions  such as solitons and Bernstein-Greene-Kruskal   and Keen modes. Results of  different gyrokinetic numerical investigations were presented, mostly in connection with the physics of  magnetized fusion plasmas.
In the context of relativistic laser produced plasmas and particle beam generation the problems of the plasma wake field dynamics, of plasma particle acceleration and   of  the wavebreak in a thermal plasma were addressed.
A major subject tackled   in this topical issue  is  the phenomenology of plasma turbulence and  intermittence  and in particular of  phase space turbulence  and its investigation within the Vlasov  or the gyrokinetic frameworks for both laboratory and space plasmas, but the most numerous set of  articles addresses numerical problems and schemes in the numerical  integration of the Vlasov-Poisson or Vlasov-Maxwell   systems both for  magnetized plasmas  and for laser produced relativistic plasmas and particle acceleration.  
Plasma sheaths  and low temperature and  dusty plasmas are  also discussed, the latter in one article as contaminants of a pure electron plasma.
On  less  traditional aspects,  we can mention  the role of the Vlasov equation in the study of nuclear collective dynamics,  the inclusion of exchange interactions in  the Vlasov equation for a spin plasma, a cosmological application of an axionic extension of the Maxwell-Vlasov theory and finally  the derivation of  a Vlasov  equation not for the particles in the plasma but for photons and quasi-particles  associated with wave modes  leading to the definition of a quasi-particle susceptibility.

\section{Conclusions and perspectives}

Two major points are evident from  reading the articles collected  in this topical issue. \\
1) The level of theoretical and numerical maturity that has been reached by the investigation of the collective dynamics of nearly dissipationless   electromagnetic  or   gravitational multi-particle systems: it is now becoming possible  to simulate directly  within a well defined mathematical framework and without unrealistic  parameter restrictions,  fully nonlinear kinetic regimes in $6 + 1 $-dimensions (six dimensions in phase space plus time). This has enormous implications from cosmology to fusion science, also at the purely conceptual level where the use of simplified models can easily end up playing a ``conservative'' role. \\
2) The  extended scope that the Vlasov equation can have in fields beyond its standard range of applications, including e.g. the novel field of collective Quantum Electrodynamics in the context of high energy density plasmas.
 
% Non-BibTeX users please use

\end{document}